\documentclass[12pt]{iopart}
\usepackage{iopams}
\usepackage{dsfont}
\usepackage[dvips]{graphicx}
\newcommand{\half}{\mbox{$\textstyle \frac{1}{2}$}}

\newcommand{\re}{\mbox{$\rm e$}}
\newcommand{\ri}{\mbox{$\rm i$}}
\newcommand{\rd}{\mbox{$\rm d$}}

\begin{document}

\title[Metric approach to quantum constraints]
{Metric approach to quantum constraints}

\author[D~C~Brody, A~C~T~Gustavsson, and
L~P~Hughston]{Dorje~C~Brody${}^1$, Anna~C~T~Gustavsson${}^2$, and
Lane~P~Hughston${}^1$}

\address{${}^1$ Department of Mathematics, Imperial College London,
London SW7 2AZ, UK}

\address{${}^2$ Blackett Laboratory,
Imperial College London, London SW7 2AZ, UK}

\begin{abstract}
A new framework for deriving equations of motion for constrained
quantum systems is introduced, and a procedure for its
implementation is outlined. In special cases the framework reduces
to a quantum analogue of the Dirac theory of constrains in classical
mechanics. Explicit examples involving spin-$\frac{1}{2}$ particles
are worked out in detail: in one example our approach coincides with
a quantum version of the Dirac formalism, while the other example
illustrates how a situation that cannot be treated by Dirac's
approach can nevertheless be dealt with in the present scheme. \\
\end{abstract}

\submitto{\JPA} \vspace{0.4cm}

\section{Introduction}

Recently there has been a renewed interest in understanding the
properties of constrained quantum dynamics \cite{Buric,Corichi,GKN,
BGH-CQM1}. The key idea behind quantum constraints is the fact that
the space of pure states (rays through the origin of Hilbert space)
is a symplectic manifold, and hence that Dirac's theory of
constraints \cite{Dirac-50,Dirac-58} in classical mechanics is
applicable in the quantum regime. The quantum state space is also
equipped with a metric structure---generally absent in a classical
phase space---induced by the probabilistic features of quantum
mechanics. In the context of analysing constrained quantum motions
it is therefore natural to examine the theory from the viewpoint of
metric geometry, as opposed to a treatment based entirely on
symplectic geometry. This is the goal of the present paper.

The metric approach that we propose is not merely a reformulation of
the Dirac formalism using the quantum symplectic structure. Indeed,
there are two distinct advantages in the metric approach over the
symplectic approach: (a) the metric approach to quantum constraints
is in general more straightforward to implement, even in situations
when the constraints can be treated by the symplectic method; and
(b) there are nontrivial examples of constraints that cannot be
implemented in the symplectic approach but can be implemented in the
metric approach. Our plan therefore is first to outline the general
metric approach to quantum constraints, and then to consider
specific examples. We also derive and examine a necessary condition
for the metric approach to be equivalent to the symplectic
formalism. The first example examined here concerns the system
consisting of a pair of spin-$\frac{1}{2}$ particles. We impose the
constraint that the state should lie on the product subspace of the
total state space, upon which all the energy eigenstates lie. This
is the example considered in \cite{Buric,BGH-CQM1} using the
symplectic approach. Here we analyse the problem using the metric
approach, and show that the constrained equations of motion reduce
to those obtained in \cite{BGH-CQM1}. The second example concerns a
single spin-$\frac{1}{2}$ particle, and we impose the constraint
that an observable that does not commute with the Hamiltonian must
be conserved. This is perhaps the simplest example of a quantum
constraint that is not evidently tractable in the symplectic
approach but can be readily dealt with by use of the metric
approach.

\section{Geometry of quantum state space}

We begin by remarking that the space of pure quantum states
associated with a Hilbert space of dimension $n$ is the projective
Hilbert space ${\mathcal P}^{n-1}$ of dimension $n-1$ (see
\cite{Kibble,cirelli, Hughston,ashtekar,gqm} and references cited
therein). We regard ${\mathcal P}^{n-1}$ as a real even-dimensional
manifold $\Gamma$, and denote a typical point in $\Gamma$,
corresponding to a ray in the associated Hilbert space, by
$\{x^a\}_{a=1,2,\ldots, 2n-2}$. It is well known that $\Gamma$ has
an integrable complex structure. Since the complex structure of
$\Gamma$ plays an important role in what follows, it may be helpful
if we make a few general remarks about the relevant ideas.

We recall that an even-dimensional real manifold ${\mathfrak M}$ is
said to have an almost complex structure if there exists a global
tensor field $J^a_{b}$ satisfying
\begin{eqnarray}
J^a_{c} J^c_{b} = -\delta^a_{b}, \label{JJdelta}
\end{eqnarray}
The almost complex structure is then said to be integrable if the
Nijenhuis tensor
\begin{eqnarray}
N_{ab}^c = J^c_{d} \nabla_{[a} J^d_{b]} - J^d_{[a} \nabla_{|d|}
J^c_{b]}
\end{eqnarray}
vanishes \cite{NN}. It is straightforward to check that $N_{ab}^c$
is independent of the choice of symmetric connection $\nabla_a$ on
${\mathfrak M}$. The vanishing of $N_{ab}^c$ can be interpreted as
follows. A complex vector field on ${\mathfrak M}$ is said to be of
positive (\textit{resp}., negative) type if $J^a_{b} V^b = +\ri V^a$
(\textit{resp}., $J^a_{b} V^b = -\ri V^a$). The vanishing of
$N_{ab}^c$ is a necessary and sufficient condition for the
commutator of two vector fields of the same type to be of that type.

A Riemannian metric $g_{ab}$ on ${\mathfrak M}$ is said to be
compatible with an almost complex structure $J^a_{b}$ if the
following conditions hold: (i) the metric is Hermitian:
\begin{eqnarray}
J^a_{c}J^b_{d} g_{ab}=g_{cd}, \label{gOmegaJ}
\end{eqnarray}
and (ii) the almost complex structure is covariantly constant:
\begin{eqnarray}
\nabla_a J^b_c = 0, \label{eq:2}
\end{eqnarray}
where $\nabla_a$ is the torsion-free Riemannian connection
associated with $g_{ab}$. An alternative expression for the
Hermitian condition is that $\Omega_{ab}=-\Omega_{ba}$, where
\begin{eqnarray}
\Omega_{ab}=J^c_{a}g_{bc}. \label{eq:4}
\end{eqnarray}
It follows that $\nabla_a J^b_{c}=0$ if and only if $\nabla_a
\Omega_{bc}=0$. However, if the almost complex structure is
integrable, then a sufficient condition for $\nabla_a J^b_{c}=0$ is
$\nabla_{[a}\Omega_{bc]}=0$. This follows on account of the identity
\begin{eqnarray}
\nabla_a \Omega_{bc} = {\textstyle\frac{3}{2}} J^p_b J^q_c
\nabla_{[a}\Omega_{pq]} - {\textstyle\frac{3}{2}}
\nabla_{[a}\Omega_{bc]} + {\textstyle\frac{1}{4}} \Omega_{ad}
N^d_{bc}.
\end{eqnarray}
A manifold ${\mathfrak M}$ with an integrable complex structure and
a compatible Riemannian structure is called a K\"ahler manifold. The
antisymmetric tensor $\Omega_{ab}$ is then referred to as the
`fundamental two-form' or K\"ahler two-form. It follows from the
definition of $\Omega_{ab}$ along with the Hermitian condition on
$g_{ab}$ that $\Omega_{ab}$ itself is Hermitian in the sense that
\begin{eqnarray}
J^a_{c}J^b_{d} \Omega_{ab}=\Omega_{cd}. \label{gOmegaJ2}
\end{eqnarray}
Furthermore, we find the tensor $\Omega^{ab}$ defined by using the
inverse metric to raise the indices of $\Omega_{ab}$, so
\begin{eqnarray}
\Omega^{ab} = g^{ac}g^{db} \Omega_{cd},
\end{eqnarray}
acts as an inverse to $\Omega_{ab}$. In particular, we have
\begin{eqnarray}
\Omega^{ac}\Omega_{bc}=\delta^a_b.
\end{eqnarray}

In the case of quantum theory there is a natural Riemannian
structure on the manifold $\Gamma$, called the Fubini-Study metric.
If $x$ and $y$ represent a pair of points in $\Gamma$, and
$|\psi(x)\rangle$ and $|\psi(y)\rangle$ are representative Hilbert
space vectors, then the Fubini-Study distance between $x$ and $y$ is
given by $\theta$, where
\begin{eqnarray}
\frac{\langle\psi(y)|\psi(x)\rangle\langle\psi(x)|\psi(y)\rangle}
{\langle\psi(x)|\psi(x)\rangle\langle\psi(y)|\psi(y)\rangle} = \half
(1+\cos\theta).
\end{eqnarray}
The K\"ahler form $\Omega_{ab}$ can be used to define a
one-parameter family of symplectic structures on $\Gamma$, given by
$\kappa\Omega_{ab}$, where $\kappa$ is a nonvanishing real constant.
In quantum mechanics the symplectic structure defined by
\begin{eqnarray}
\omega_{ab} = \half \Omega_{ab} \label{eq:77}
\end{eqnarray}
plays a special role. In particular, if we define the inverse
symplectic structure by $\omega^{ab}=2\Omega^{ab}$ so that
$\omega^{ac}\omega_{bc}=\delta^a_b$, and if we choose units such
that $\hbar=1$, then we find that the Schr\"odinger trajectories on
$\Gamma$ are given by Hamiltonian vector fields of the form
\begin{eqnarray}
\dot{x}^a = \omega^{ab}\nabla_b H, \label{xdot}
\end{eqnarray}
where
\begin{eqnarray}
H(x)=\frac{\langle\psi(x)|{\hat H}|\psi(x)\rangle}{\langle\psi(x)|
\psi(x)\rangle}. \label{H}
\end{eqnarray}
Thus we see that the expectation of the Hamiltonian operator ${\hat
H}$ gives rise to a real function $H(x)$ on $\Gamma$. This function
plays the role of the Hamiltonian in the determination of the
symplectic flow associated with the Schr\"odinger trajectory.

\section{Metric formalism for quantum constraints}

With these geometric tools in hand we now proceed to formulate the
metric approach to quantum constraints. We assume that the system
under investigation is subject to a family of $N$ constraints
\begin{eqnarray}
\Phi^{i}(x) = 0, \label{eq:6}
\end{eqnarray}
where $i=1,\ldots,N$. The condition $\Phi^i (x)=0$ for each $i$
defines a hypersurface in $\Gamma$. The intersection of all $N$ such
hypersurfaces then defines the constraint subspace in $\Gamma$ onto
which the dynamics must be restricted. To enforce the constraint in
the metric approach we remove from the vector field ${\dot x}^a$
those components that are normal to the constraint subspace. The
equations of motion then take the form
\begin{eqnarray}
\dot{x}^a = \omega^{ab}\nabla_b H - \lambda_{i} g^{ab}\nabla_b
\Phi^{i}, \label{EOM}
\end{eqnarray}
where $\omega^{ab}$ is the inverse quantum symplectic structure, and
where $\{\lambda_{i}\}_{i=1,\ldots,N}$ are the Lagrange multipliers
associated with the constraints. In order to determine
$\{\lambda_{i}\}$ we consider the identity $\dot{\Phi}^{j} = 0$.
Using the chain rule this can be expressed as
\begin{eqnarray}
\dot{x}^a \nabla_a \Phi^{j} = 0. \label{EOM2}
\end{eqnarray}
Substitution of (\ref{EOM}) in (\ref{EOM2}) then gives
\begin{eqnarray}
\omega^{ab} \nabla_a \Phi^{j} \nabla_b H - \lambda_{i} g^{ab}
\nabla_a \Phi^{j} \nabla_b \Phi^{i}  = 0 \label{eqn7}.
\end{eqnarray}
To simplify (\ref{eqn7}) we define the symmetric matrix
\begin{eqnarray}
M^{ij}=g^{ab} \nabla_a \Phi^{i} \nabla_b \Phi^{j}. \label{eq:9}
\end{eqnarray}
If the matrix $M^{ij}$ is nonsingular, we write $M_{ij}$ for its
inverse such that $M_{ik} M^{kj}=\delta^{j}_{i}$. We can then solve
(\ref{eqn7}) for $\lambda_i$ to find
\begin{eqnarray}
\lambda_{i}=M_{ij} \omega^{ab} \nabla_a \Phi^{j} \nabla_b H.
\label{eq:11}
\end{eqnarray}
Substituting the expression for $\lambda_{i}$ back into
(\ref{EOM}) we find that the constrained equations of motion are
given by
\begin{eqnarray}
\dot{x}^a = \omega^{ab} \nabla_b H - M_{ij} \omega^{cd}\nabla_c
\Phi^{j} \nabla_d H g^{ab} \nabla_b \Phi^{i}. \label{eq:12}
\end{eqnarray}
This is the main result of the paper.

We remark that in the case for which the system of constraints
$\{\Phi^{i}\}$ corresponds to a family of quantum observables,
$M^{ij}$ for fixed $i,j$ represents the covariance between the two
observables $\hat{\Phi}^{i}$ and $\hat{\Phi}^{j}$ in the state
represented by the point $x$:
\begin{eqnarray}
M^{ij} = \langle \hat{\Phi}^i \hat{\Phi}^j\rangle - \langle
\hat{\Phi}^i \rangle\langle \hat{\Phi}^j\rangle. \label{M2}
\end{eqnarray}
It follows that if there is a single constraint for a conserved
observable (as in Example~2 below), then $M$ is never singular
except at isolated points of zero measure because the variance of an
observable is positive except at its eigenstates. In the case of two
conserved observables, say, $\hat{\Phi}^1=\hat{A}$ and
$\hat{\Phi}^2=\hat{B}$, we have
\begin{eqnarray}
M^{ij} = \left( \begin{array}{cc} {\rm var}(\hat{A}) &
{\rm cov}(\hat{A},\hat{B}) \\
{\rm cov}(\hat{A},\hat{B}) & {\rm var}(\hat{B}) \end{array} \right)
.
\end{eqnarray}
The determinant $\Delta$ of $M^{ij}$ is then given by
\begin{eqnarray}
\Delta = (1-\rho^2)\,{\rm var}(\hat{A})\,{\rm var}(\hat{B}),
\end{eqnarray}
where $\rho$ is the correlation of $\hat{A}$ and $\hat{B}$. Since
the variances are positive (except at eigenstates), $\Delta$ can
identically vanish if and only if $\rho=\pm1$, i.e. either $\hat{A}$
and $\hat{B}$ are perfectly correlated or they are perfectly
anticorrelated. Analogous observations can be made when there are
more than two conserved observables for the constraint, and we
conclude that $M^{ij}$ is singular only when the constraints are
redundant. Therefore, in the case of constraints on conserved
observables the metric approach introduced here is always
implementable, provided that we do not introduce redundant
constraints.

In the case of algebraic constraints for which the constraint system
$\{\Phi^{i}\}$ does not correspond to a family of quantum
observables (as in Example~1 below), we are unaware of any physical
characteristic that might lead to a singular behaviour in $M^{ij}$.
Thus in this case the invertibility of $M^{ij}$ must be examined
individually.

\section{Equivalence of metric and symplectic approaches}

Before considering specific examples of constrained systems that can
be described using the present approach, it will be of interest to
ask how this framework might be related to the approach of Dirac
\cite{Dirac-50,Dirac-58}, or more precisely, its quantum counterpart
\cite{Buric,Corichi,BGH-CQM1} which we shall refer to as the
symplectic approach. In the symplectic approach, the constrained
equations of motion can be expressed in the same form as
(\ref{xdot}), but with a modified inverse symplectic structure
$\tilde{\omega}^{ab}$ in place of $\omega^{ab}$, which in effect is
the induced symplectic form on the constraint surface
\cite{BGH-CQM1}. Thus, we would like to know under what condition
the metric approach leading to the right side of (\ref{eq:12}) also
reduces to a modified symplectic flow of the form
$\tilde{\omega}^{ab}\nabla_b H$, where $H$ is the same Hamiltonian
as the one in the original Schr\"odinger equation (\ref{xdot}).
Intuitively, we would expect that when there is an even number of
constraints, the two methods might become equivalent. However, the
verification or falsification of this assertion turns out to be
difficult. Nevertheless, in what follows we shall establish the
sufficient condition under which the symplectic approach and the
metric approach become equivalent.

In order to investigate whether (\ref{eq:12}) can be rewritten in
the form
\begin{eqnarray}
\dot{x}^a = \tilde{\omega}^{ab} \nabla_b H \label{CEOM2}
\end{eqnarray}
for a suitably defined antisymmetric tensor $\tilde{\omega}^{ab}$,
we rearrange terms in (\ref{eq:12}) to write
\begin{eqnarray}
\dot{x}^a &=& \left(\omega^{ad}  - M_{ij} g^{ab} \nabla_b \Phi^i
\omega^{cd}\nabla_c \Phi^j \right)\nabla_d H \nonumber \\ &=&
\left(\omega^{ad}-\mu_{bc} g^{ab} \omega^{cd}\right)\nabla_d H,
\end{eqnarray}
where we have defined
\begin{eqnarray}
\mu_{bc}=M_{ij} \nabla_b \Phi^i\nabla_c \Phi^j. \label{mu}
\end{eqnarray}
It follows that we need to find whether the expression
\begin{eqnarray}
\tilde{\omega}^{ab} = \omega^{ab}-g^{ad} \omega^{cb}\mu_{dc}
\end{eqnarray}
defines a symplectic structure on the subspace of the state space.
Since the symplectic structure $\omega^{ab}$ is antisymmetric, we
shall examine under which condition $\tilde{\omega}^{ab}$ is
antisymmetric. This is equivalent to asking whether the following
relation holds:
\begin{eqnarray}
g^{ac}\omega^{bd}\mu_{cd} \stackrel{?}{=} - g^{bc} \omega^{ad}
\mu_{cd}.\label{assump1}
\end{eqnarray}
Suppose that (\ref{assump1}) is valid. Then using (\ref{eq:4}) and
(\ref{eq:77}) we can rewrite (\ref{assump1}) as
\begin{eqnarray}
g^{ac}g^{be}J^d_e \mu_{cd} = - g^{bc}g^{ae}J^d_e \mu_{cd}.
\end{eqnarray}
Transvecting this with $g_{fa}$ and relabelling the indices we
obtain
\begin{eqnarray}
g^{be}J^d_e \mu_{ad} = - g^{bc}J^d_a \mu_{cd},
\end{eqnarray}
from which it follows that
\begin{eqnarray}
J^d_b \mu_{ad} = - J^d_a \mu_{bd}. \label{assump2}
\end{eqnarray}
Multiplying both sides of (\ref{assump2}) with $J^a_c$ we find that
the condition (\ref{assump1}) is equivalent to
\begin{eqnarray}
J^c_a J^d_b \mu_{cd} = \mu_{ab}, \label{cond1}
\end{eqnarray}
where we have used (\ref{JJdelta}) and the symmetry of $\mu_{ab}$.
It follows that for the metric formulation of the constraint motion
(\ref{eq:12}) to be expressible in the Hamiltonian form
(\ref{CEOM2}) with the original Hamiltonian $H$ the matrix
$\mu_{bc}$ defined by (\ref{mu}) must be Hermitian. This is the
sufficiency condition that we set out to establish. Let us now
examine specific cases to gain insights into this condition.
\vspace{0.4cm}

\emph{Single constraint case.} In general the $J$-invariance
condition (\ref{cond1}) need not be satisfied. One can easily see
this by considering the case for which there is only one constraint
given by $\Phi(x)=0$. Then expression (\ref{eq:9}) becomes a scalar
quantity, which we will denote by $M$, and thus its inverse is
$M^{-1}$. It follows that
\begin{eqnarray}
\mu_{ab}=M^{-1}\nabla_a \Phi \nabla_b \Phi.
\end{eqnarray}
Substituting this expression for $\mu_{ab}$ into (\ref{cond1}) gives
us
\begin{eqnarray}
J^c_a\nabla_c \Phi J^d_b \nabla_d \Phi = \nabla_a \Phi \nabla_b
\Phi,
\end{eqnarray}
which implies that
\begin{eqnarray}
J^c_a\nabla_c \Phi =\nabla_a \Phi,
\end{eqnarray}
but this clearly is a contradiction, since the two vectors
$J^c_a\nabla_c \Phi$ and $\nabla_a \Phi$ are \textit{orthogonal}. We
see therefore that in the case of a single constraint the condition
(\ref{cond1}) is not satisfied, and the constraint motion
(\ref{eq:12}) cannot be expressed in the Hamiltonian form
(\ref{CEOM2}). \vspace{0.4cm}

\emph{Two constraint case.} Let us examine the case in which there
are two constraints. As we shall show, in this case the
$J$-invariance condition for $\mu_{bc}$ reduces to a simpler
condition. Let us write $\Phi^1=A$ and $\Phi^2=B$ for the two
constraints. Then the inverse of the matrix $M^{ij}$ can be written
in the form
\begin{eqnarray}
M_{ij} = \frac{1}{2\Delta}\, \epsilon_{ii'}\epsilon_{jj'} M^{i'j'},
\label{M_ij}
\end{eqnarray}
where $\Delta=\textrm{det}(M^{ij})$, and $\epsilon_{ij}$ is a
totally-skew tensor with $i,i',j,j'=1,2$. Substituting (\ref{M_ij})
into (\ref{mu}) we find
\begin{eqnarray}
\mu_{bc} &=& M_{ij}\nabla_b \Phi^i \nabla_c \Phi^j \nonumber\\
&=& \frac{1}{2\Delta}\,\epsilon_{ii'}\epsilon_{jj'}g^{pq} \nabla_p
\Phi^{i'} \nabla_q \Phi^{j'} \nabla_b \Phi^i\nabla_c \Phi^j
\nonumber\\ &=&\frac{1}{2\Delta}\,\epsilon_{ii'}\nabla_b
\Phi^i\nabla_p \Phi^{i'} \epsilon_{jj'}\nabla_c \Phi^j \nabla_q
\Phi^{j'} g^{pq} \nonumber\\ &=&
\frac{1}{2\Delta}\,\tau_{bp}\tau_{cq}g^{pq}, \label{mu-2const}
\end{eqnarray}
where we have defined
\begin{eqnarray}
\tau_{ab}=\epsilon_{ij}\nabla_a \Phi^i \nabla_b \Phi^j.
\end{eqnarray}
In order for (\ref{mu-2const}) to satisfy (\ref{cond1}) we thus
require that
\begin{eqnarray}
J^b_a \tau_{bc} = \pm \tau_{ab} J^b_c, \label{cond2}
\end{eqnarray}
where we have substituted (\ref{mu-2const}) into condition
(\ref{cond1}) and used $J^b_p J^c_q g^{pq} = g^{bc}$. Since
\begin{eqnarray}
\tau_{ab} = \nabla_a A \nabla_b B - \nabla_a B \nabla_b A,
\label{eqn32}
\end{eqnarray}
we find that (\ref{cond2}) can be written more explicitly in the
form
\begin{eqnarray}
J^b_a\nabla_b A \nabla_c B - J^b_a \nabla_b B \nabla_c A = \pm\left(
\nabla_a A J^b_c \nabla_b B - \nabla_a B J^b_c\nabla_b A\right).
\label{eqn33}
\end{eqnarray}
Hence if this condition is satisfied (with either a plus or a minus
sign) by the two constraints $A$ and $B$, then the constrained
equations of motion (\ref{eq:12}) take the form (\ref{CEOM2}).
\vspace{0.4cm}

\emph{Holomorphic constraints.} If we choose the constraint function
to be holomorphic so that the two constraints $A$ and $B$ are given
by the real and imaginary parts of
\begin{eqnarray}
\Phi(x)=A(x)+\ri B(x), \label{PhiAB}
\end{eqnarray}
then we can find the form of $\tau_{ab}$ for which (\ref{cond2})
holds. To obtain an expression for $\tau_{ab}$, we recall that a
real vector on $\Gamma$ can be decomposed into its complex
`positive' and `negative' parts $V_a=(V_\alpha,V_{\alpha'})$ (cf.
\cite{bh98}). These components are given respectively by
\begin{eqnarray}
(V_\alpha,0) =  \half(V_a - \ri J^b_a V_b), \quad {\rm and} \quad
(0,V_{\alpha'}) = \half(V_a + \ri J^b_a V_b).
\end{eqnarray}
Hence, $V_{\alpha'}$ is the complex conjugate of $V_\alpha$, and
these components are the eigenvectors of the complex structure
$J^a_b$ with eigenvalues $\pm\ri$. It follows that if the vector is
of type $V_a=(V_\alpha,0)$ then $J^a_b V_a=\ri V_b$, and similarly
if $V_a=(0,V_{\alpha'})$ then $J^a_b V_a = -\ri V_b$. With respect
to this decomposition the tensor $\tau_{ab}$ can be expressed as
\begin{eqnarray}
\tau_{ab}=\left( \begin{array}{cc} \tau_{\alpha\beta} &
\tau_{\alpha\beta'} \\ \tau_{\alpha'\beta} & \tau_{\alpha'\beta'}
\end{array} \right). \label{tau-decomp}
\end{eqnarray}
We rewrite the condition (\ref{cond2}) in the form
\begin{eqnarray}
J^c_a J^d_b \tau_{cd}=\pm \tau_{ab} \label{cond3}
\end{eqnarray}
by contracting both sides with $J^c_d$ and using (\ref{JJdelta}).
In terms of the decomposition (\ref{tau-decomp}) we find that in
order for (\ref{cond3}) to be true we require that $\tau_{ab}$ takes
either of the two forms:
\begin{eqnarray}
\tau^{(+)}_{ab}=\left( \begin{array}{cc} 0 & \tau_{\alpha\beta'}
\\ \tau_{\alpha'\beta} & 0 \end{array} \right), \qquad \textrm{or}
\qquad \tau^{(-)}_{ab}=\left( \begin{array}{cc} \tau_{\alpha\beta} &
0 \\ 0 & \tau_{\alpha'\beta'} \end{array} \right),
\label{tau-plusminus}
\end{eqnarray}
where the plus and minus in $\tau^{(\pm)}_{ab}$ correspond to the
required sign in (\ref{cond3}). In view of (\ref{PhiAB}) we can
write the two constraints as $A(x)=\half(\Phi+\bar{\Phi})$ and
$B(x)=-\half\ri(\Phi-\bar{\Phi})$, where $\bar{\Phi}$ denotes the
complex conjugate of $\Phi$. Then we have:
\begin{eqnarray}
\begin{array}{ll} \nabla_\alpha A = \half\nabla_\alpha \Phi,
&  \qquad \nabla_\alpha B  = -\half\ri \nabla_\alpha \Phi, \\
\nabla_{\alpha'} A  = \half \nabla_{\alpha'}\bar{\Phi}, & \qquad
\nabla_{\alpha'} B  = \half \ri \nabla_{\alpha'}\bar{\Phi}.
\end{array}
\end{eqnarray}
Using these expressions together with (\ref{eqn32}) we find that the
components of $\tau_{ab}^{(-)}$ all vanish. Hence, when the two
constraints $A$ and $B$ are given by (\ref{PhiAB}), the condition
(\ref{cond2}) is satisfied when $\tau_{ab}$ takes the form
\begin{eqnarray}
\tau_{ab}=\left( \begin{array}{cc} 0 & \half\ri\nabla_\alpha \Phi
\nabla_{\beta'}\bar{\Phi} \\ -\half\ri \nabla_{\alpha'}\bar{\Phi}
\nabla_\beta \Phi & 0 \end{array} \right), \label{finaltau}
\end{eqnarray}
i.e. when $\tau_{ab}=\tau_{ab}^{(+)}$. It follows that the metric
formalism associated with holomorphic constraints of the form
(\ref{PhiAB}) is equivalent to the symplectic formalism.

We remark that the quadratic form $\tilde{\omega}^{ab}$ acting from
the right annihilates the vector $\nabla_a\Phi^k$ normal to the
constraint surface. This can be verified explicitly as follows:
\begin{eqnarray}
\tilde{\omega}^{ad}\nabla_a \Phi^k & = & \omega^{ad}\nabla_a \Phi^k
- M_{ij} g^{ab}\nabla_b \Phi^i \nabla_a \Phi^k \omega^{cd} \nabla_c
\Phi^j \nonumber \\ & = & \omega^{ad}\nabla_a \Phi^k - M_{ij} M^{ki}
\omega^{cd}\nabla_c \Phi^j \nonumber \\ &=& \omega^{ad}\nabla_a
\Phi^k - \omega^{ad} \nabla_a \Phi^k \nonumber \\ &=&0 ,
\end{eqnarray}
since $g^{ab}\nabla_a\Phi^k \nabla_b \Phi^i = M^{ki}$ and
$M_{ij}M^{ki}=\delta^k_j$. This condition, of course, is equivalent
to the condition that ${\dot x}^a\nabla_a \Phi^k=0$. However, in
general we have
\begin{eqnarray}
\tilde{\omega}^{ad}\nabla_d \Phi^k \neq 0, \label{eq:xx30}
\end{eqnarray}
since $\tilde{\omega}^{ab}$ is not in general antisymmetric. In
other words, the vanishing of the left side of (\ref{eq:xx30}) is
equivalent to the $J$-invariance condition for $\mu_{bc}$.

In summary, the procedure for deriving the equations of motion in
the metric approach to constrained quantum motion is as follows.
First, express the relevant constraints in the form
$\{\Phi^i(x)=0\}_{i=1,2,\ldots,N}$. Determine the matrix $M^{ij}$
via (\ref{eq:9}). Assuming that $M^{ij}$ is nonsingular, calculate
its inverse $M_{ij}$. Substitute the result in (\ref{eq:12}), and we
recover the relevant equations of motion. Having obtained the
general procedure, let us now examine some explicit examples
implementing this procedure.

\section{Illustrative examples}

{\em Example 1}. The first example that we consider here is
identical to the one considered in \cite{BGH-CQM1} involving a pair of
spin-$\frac{1}{2}$ particles. We consider the subspace of the state
space associated with product states upon which all the energy
eigenstates lie. An initial state that lies on this product space is
required to remain a product state under the evolution generated by
a generic Hamiltonian. In this example there are two constraints
$\Phi^1(x)$ and $\Phi^2(x)$, and we shall show that the metric
approach introduced here gives rise to a result that agrees with the
one obtained in \cite{BGH-CQM1} using the symplectic approach.

Let us work with the coordinates of the quantum state space given by
the `action-angle' variables \cite{BGH-CQM1,oh}, where the canonical
conjugate variables are given by $\{x^a\} =\{q_{\nu},
p_{\nu}\}_{\nu=1,\ldots,n-1}$ such that when a generic pure states
$|x\rangle$ is expanded in terms of the energy eigenstates $\{|
E_{\alpha}\rangle\}_{\alpha=1,\ldots,n}$, the associated amplitudes
are given by $\{p_\nu\}$, and the relative phases by
$\{q_\nu\}$. In the case of a pair of spin-$\half$ particles we can
thus expand a generic state in the form
\begin{eqnarray}
\fl |x\rangle = \sqrt{p_1} \re^{-{\rm i} q_1} |E_1\rangle +
\sqrt{p_2} \re^{-{\rm i} q_2} |E_2\rangle + \sqrt{p_3} \re^{-{\rm i}
q_3} |E_3\rangle + \sqrt{1-p_1-p_2-p_3} |E_4\rangle. \label{psi}
\end{eqnarray}
This choice of coordinates has the property that if we write
\begin{eqnarray}
H(p,q) = \frac{\langle x | H | x \rangle}{\langle x|x\rangle}
\end{eqnarray}
for the Hamiltonian, where $|x\rangle=|p,q\rangle$, then the
Schr\"odinger equation is expressed in the form of the conventional
Hamilton equations:
\begin{eqnarray}
\dot{q}_{\nu}=\frac{\partial H(q,p)}{\partial p_{\nu}} \quad {\rm
and} \quad \dot{p}_{\nu}=-\frac{\partial H(q,p)}{\partial q_{\nu}}.
\end{eqnarray}
In the energy basis the Hamiltonian can be expressed as
\begin{eqnarray}
H=\sum_{\alpha=1}^4 E_{\alpha} |E_{\alpha}\rangle\langle
E_{\alpha}|.
\end{eqnarray}
It follows that the phase space function $H(p,q)$ for a generic
Hamiltonian is given by
\begin{eqnarray}
H=E_4 + \sum_{{\nu}=1}^3 \Omega_{\nu} p_{\nu},
\end{eqnarray}
where $\Omega_{\nu} = E_{\nu} - E_4$. The symplectic structure and
its inverse in these coordinate are thus given respectively by
\begin{eqnarray}
\omega_{ab} = \left(\begin{array}{cc}{\mathds O} & {\mathds 1} \\
-{\mathds 1} & {\mathds O} \end{array} \right) \quad \textrm{and}
\quad \omega^{ab} = \left(\begin{array}{cc}{\mathds O} & {\mathds 1}
\\ -{\mathds 1} & {\mathds O} \end{array} \right), \label{symp}
\end{eqnarray}
where ${\mathds 1}$ is the $2\times 2$ identity matrix and ${\mathds
O}$ is the $2\times 2$ null matrix.

Let us write $\{\psi^\alpha\}_{\alpha=1,2,3,4}$ for the coefficients
of the energy eigenstates in (\ref{psi}). Then the constraint
equation is expressed in the form $\psi^1\psi^4=\psi^2\psi^3$, which
is just a single complex equation \cite{BGH-CQM1}. Thus in real
terms we have two constraints given by
\begin{eqnarray}
\left\{ \begin{array}{l} \Phi^1 =\sqrt{p_1 p_4} \cos q_1 - \sqrt{p_2
p_3} \cos (q_2 + q_3) = 0, \\ \Phi^2 =  \sqrt{p_1 p_4} \sin q_1 -
\sqrt{p_2 p_3} \sin (q_2 + q_3) = 0. \end{array} \right.
\end{eqnarray}
These constraints are `separable' and can be rewritten as
\begin{eqnarray}
\left\{ \begin{array}{l} \Phi^1 = q_1 - q_2 - q_3 = 0, \\
\Phi^2 = p_1 (1-p_1-p_2-p_3) - p_2 p_3 = 0. \end{array} \right.
\end{eqnarray}

The next step in the metric approach for the constraint is to work
out the expression for the Fubini-Study metric associated with the
line element:
\begin{eqnarray}
\rd s^2 = 8\left[\frac{\psi^{[\alpha}\rd \psi^{\beta]}
\bar{\psi}_{[\alpha}\rd\bar{\psi}_{\beta]}}{(\bar{\psi}_{\gamma}
\psi^{\gamma})^2} \right]
\end{eqnarray}
in terms of the canonical variables $\{q_{\nu},p_{\nu}
\}_{{\nu}=1,2,3}$. A calculation shows that
\begin{eqnarray}
\fl g_{ab} = \left(\begin{array}{cccccc}
4(1-p_1) p_1 & -4p_1 p_2 & -4 p_1 p_3&0&0&0 \\
-4p_1 p_2 & 4 ( 1-p_2) p_2 & -4 p_2 p_3 &0&0&0 \\
-4p_1 p_3 & -4p_2 p_3 & 4 ( 1-p_3) p_3 &0&0&0 \\
0&0&0& \frac{1-p_2-p_3}{p_1 p_4} & \frac{1}{p_4}&\frac{1}{p_4} \\
0&0&0& \frac{1}{p_4} & \frac{1-p_1-p_3}{p_2 p_4}&\frac{1}{p_4} \\
0&0&0& \frac{1}{p_4} & \frac{1}{p_4}&\frac{1-p_1-p_2}{p_3 p_4}
\end{array} \right),
\end{eqnarray}
where for simplicity we have denoted $p_4=1-p_1-p_2-p_3$. The
inverse is thus:
\begin{eqnarray}
\fl g^{ab} = \left(\begin{array}{cccccc}
\frac{1-p_2-p_3}{4p_1 p_4} & \frac{1}{4p_4}&\frac{1}{4p_4}&0&0&0 \\
\frac{1}{4p_4} & \frac{1-p_1-p_3}{4p_2 p_4}&\frac{1}{4p_4}&0&0&0 \\
\frac{1}{4p_4} & \frac{1}{4p_4}&\frac{1-p_1-p_2}{4p_3 p_4}&0&0&0 \\
0&0&0&(1-p_1) p_1 & -p_1 p_2 & - p_1 p_3 \\
0&0&0&-p_1 p_2 &  ( 1-p_2) p_2 & - p_2 p_3  \\
0&0&0&-p_1 p_3 & -p_2 p_3 &  ( 1-p_3) p_3  \\
\end{array} \right).
\end{eqnarray}
To obtain explicit expressions for equations of motion
(\ref{eq:12}) we need to calculate the matrix $M^{ij}$ and its
inverse. A short calculation shows that
\begin{eqnarray}
\fl M^{ij}=\left( \begin{array}{cc} \frac{1}{4}\left(\frac{1}{p_1} +
\frac{1}{p_2}+ \frac{1}{p_3}+\frac{1}{p_4} \right) & 0\\ 0 &
\begin{array}{l} p_1 p_4\left(1-4p_1p_4 + 4 p_2p_3\right) \\
+ \left(p_2+p_3-4p_2 p_3\right) \left(p_2p_3-p_1p_4 \right)
\end{array} \end{array} \right),
\end{eqnarray}
from which its inverse $M_{ij}$ can easily be obtained. Putting
these together, we find that the equations of motion are given by
\begin{eqnarray}
\begin{array}{l}
\dot{q}_1 = \Omega_1 - \frac{p_2 p_3 (1-2p_1-p_2-p_3)(\Omega_1 -
\Omega_2 - \Omega_3)}{p_2 p_3(1-p_2-p_3)-p_1^2 (p_2+p_3) +
p_1(1-p_2-p_3)(p_2+p_3)} \\ \dot{q}_2 = \Omega_2 + \frac{p_1
p_3(1-p_1-p_3) (\Omega_1-\Omega_2-\Omega_3)}{p_2
p_3(1-p_2-p_3)-p_1^2 (p_2+p_3) + p_1(1-p_2-p_3)(p_2+p_3)}\\
\dot{q}_3 = \Omega_3 + \frac{p_1 p_2(1-p_1-p_2)(\Omega_1 - \Omega_2
- \Omega_3)}{p_2
p_3(1-p_2-p_3)-p_1^2 (p_2+p_3) + p_1(1-p_2-p_3)(p_2+p_3)}\\
\dot{p_1}=0 \\ \dot{p_2}=0 \\ \dot{p_3}=0.
\end{array} \label{qdot-pdot-org}
\end{eqnarray}
We can simplify the equations using the relation $p_1 p_4 = p_2
p_3$, which gives us
\begin{eqnarray}
\begin{array}{l}
\dot{q}_1 = \Omega_1 - (1-2p_1-p_2-p_3)(\Omega_1 - \Omega_2 -
\Omega_3) \\
\dot{q}_2 = \Omega_2 + (p_1+p_3)(\Omega_1-\Omega_2-\Omega_3)\\
\dot{q}_3 = \Omega_3 + (p_1+p_2)(\Omega_1 - \Omega_2 - \Omega_3)\\
\dot{p_1}=0 \\ \dot{p_2}=0 \\ \dot{p_3}=0. \end{array}
\label{qdot-pdot}
\end{eqnarray}
These equations are precisely the ones obtained in \cite{BGH-CQM1}
by means of the symplectic formalism. The solution to these
equations are also worked out in \cite{BGH-CQM1}.

Before we proceed to the next example, let us verify explicitly that
the condition (\ref{cond1}) for the equivalence of the metric and
symplectic approaches indeed holds in this example. For this we need
to obtain the expression for the complex structure in the canonical
coordinates $\{q_{\nu},p_{\nu} \}_{{\nu}=1,2,3}$ using the relations
(\ref{eq:4}) and (\ref{eq:77}). This is given by
\begin{eqnarray}
\fl J^a_b = \left(\begin{array}{cccccc}0&0&0&
\frac{1-p_2-p_3}{2p_1p_4}& \frac{1}{2p_4}&\frac{1}{2p_4} \\
0&0&0&\frac{1}{2p_4}&\frac{1-p_1-p_3}{2p_2p_4}&\frac{1}{2p_4} \\
0&0&0&\frac{1}{2p_4}&\frac{1}{2p_4}&\frac{1-p_1-p_2}{2p_3p_4}\\
2(p_1-1)p_1 & 2p_1 p_2 & 2p_1 p_3 &0&0&0\\
2p_1 p_2 & 2(p_2-1)p_2 & 2p_2 p_3 &0&0&0\\
2p_1 p_3 & 2 p_2 p_3 & 2(p_3-1)p_3 &0&0&0 \end{array} \right),
\end{eqnarray}
where as before we write $p_4=1-p_1-p_2-p_3$. Substituting this
expression and the expression for $\mu_{ab}$ obtained from
(\ref{mu}) into (\ref{cond1}), we find that the condition is indeed
satisfied. \vspace{0.4cm}

{\em Example 2}. Consider a single spin-$\half$ particle system
immersed in a $z$-field with a unit strength. The space of pure
states for this system is just the surface of the Bloch sphere. The
Hamiltonian of the system is given by ${\hat H}={\hat \sigma}_z$,
where ${\hat \sigma}_z$ is the Pauli spin matrix in the
$z$-direction. We then impose the constraint that an observable, say
${\hat \sigma}_x$, that does not commute with the Hamiltonian, must
be conserved under the time evolution.

As before we chose the canonical coordinates $x^a=\{q,p\}$ for the
Bloch sphere by setting
\begin{eqnarray}
|x(p,q)\rangle = \sqrt{1-p}\, |E_1\rangle + \sqrt{p}\re^{-{\rm i}q}
|E_2\rangle. \label{ex2psi}
\end{eqnarray}
The Hamiltonian in this coordinate system is given by
\begin{eqnarray}
H(q,p)=1-2p.
\end{eqnarray}
The conservation of ${\hat \sigma}_x$ then reduces to a single real
constraint of the form $\langle x|{\hat \sigma}_x|x\rangle={\rm
constant}$, which, by use of (\ref{ex2psi}), gives us
\begin{eqnarray}
\Phi(x) = 2 \sqrt{p(1-p)} \cos q. \label{1particleConst}
\end{eqnarray}

\begin{figure}
\begin{center}\vspace{-0.0cm}
  \includegraphics[scale=0.5]{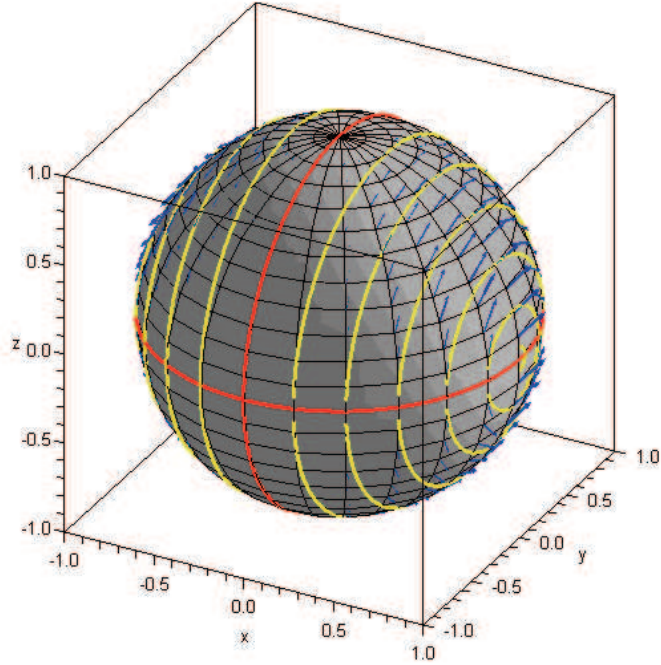}
  \vspace{-0.0cm}
  \caption{(colour online) A field plot of the dynamics
  resulting from a
  system constrained to remain on a surface defined by
  $\Phi={\hat \sigma}_x$ when the system evolves according
  to the Hamiltonian $H={\hat \sigma}_z$, where ${\hat \sigma}_x$
  and ${\hat \sigma}_z$ are Pauli matrices. Shown in
  yellow are the integral curves of the motion when we
  chose as the starting positions $\{\theta_1,\phi_n\}=
  \{\frac{23\pi}{48},\frac{2\pi n}{24}\}_{n=0,\ldots,23}$
  and $\{\theta_2,\phi_n\}=\{\frac{25\pi}{48},
  \frac{2\pi n}{24}\}_{n=0,\ldots,23}$. The great circles
  shown in red, the equators of the $x$ and $z$-axis,
  consist of fixed points. A state that initially lies
  on one of these great circles does not move away from that point,
  whereas all other states evolve asymptotically towards the fixed
  point where the associated integral curve intersects the equator
  of the sphere.
  \label{fig:3}
  }
\end{center}
\end{figure}

The metric on the Bloch sphere, in terms of our conjugate variables,
is given by
\begin{eqnarray}
g_{ab}= \left(\begin{array}{cc} 4(1-p)p & 0 \\ 0 &
\frac{1}{(1-p)p} \end{array} \right).
\end{eqnarray}
The symplectic structure has the same form as (\ref{symp}) so we can
now calculate the scalar quantity $M$ defined in (\ref{eq:9}). A
short calculation shows that
\begin{eqnarray}
M=(1-2p)^2\cos^2 q + \sin^2 q. \label{ex2m}
\end{eqnarray}
Substituting (\ref{ex2m}) into (\ref{eq:12}) we find that the
equations of motion are given by
\begin{eqnarray}
\dot{q} = \frac{-2(1-2p)^2 \cos^2 q}{(1-2p)^2\cos^2 q + \sin^2 q},
\end{eqnarray}
and
\begin{eqnarray}
\dot{p} = \frac{4(1-2p)(-1+p)p \sin q\cos q}{(1-2p)^2\cos^2 q +
\sin^2 q}.
\end{eqnarray}
In order to visualise the results we convert the equations of motion
into angular coordinates, using the method outlined in
\cite{BGH-CQM1}. In angular coordinates (\ref{ex2psi}) is given by
\begin{eqnarray}
|x(\theta,\phi)\rangle = \cos\half\theta|E_1\rangle + \sin\half
\theta\re^{\ri\phi} |E_2\rangle. \label{ex2angpsi}
\end{eqnarray}
Comparing (\ref{ex2psi}) and (\ref{ex2angpsi}) we identify $p=
\sin^2\half\theta$ and $q=-\phi$. The equations of motion then
become:
\begin{eqnarray}
\dot{\theta} = \half \left(\frac{ \sin(2\theta)\sin(2\phi)}
{1-\sin^2\theta\cos^2\phi}\right), \quad {\rm and} \quad \dot{\phi}
= \frac{2\cos^2\theta\cos^2\phi}{1-\sin^2\theta\cos^2\phi}.
\label{ex2eom2}
\end{eqnarray}
Figure \ref{fig:3} shows some of the integral curves resulting from
the above equations, plotted on the surface of the Bloch sphere.
Equation (\ref{ex2eom2}) is valid at all points except where
$(\theta, \phi)=(\frac{\pi}{2},0)$ and where $(\theta,
\phi)=(\frac{\pi}{2},\pi)$. At these singular fixed points the
method used cannot be defined.

It is also straightforward to verify that the results above could
not have been obtained using the symplectic approach. The complex
structure on the underlying state space of the spin-$\half$ particle
in our coordinate system is given by
\begin{eqnarray}
J^a_b = \left(\begin{array}{cc}0&\frac{1}{2(1-p)p}\\ -2(1-p)p & 0
\end{array} \right). \label{ex2j}
\end{eqnarray}
Evaluating (\ref{mu}) using (\ref{ex2m}), and substituting the
result along with (\ref{ex2j}) into (\ref{cond1}), we find that
(\ref{cond1}) does not hold.

\vspace{0.4cm}



\end{document}